\documentclass[trackchanges,twocolumn]{aastex701}
\usepackage{CJK}
\usepackage{amsmath,xcolor}
\usepackage{bm}
\usepackage{soul}

\begin{document}
\begin{CJK*}{UTF8}{gbsn}
\title{Pulsed Accretion onto Eccentric Binaries in Highly Misaligned Circumbinary Disks}

\author[0009-0002-5097-3559]{Ruiqi Yang (杨锐祺)}
\affiliation{School of Physics and Astronomy, Sun Yat-sen University, Zhuhai 519082, People's Republic of China}
\email[show]{yangrq9@mail2.sysu.edu.cn}

\author[0000-0002-4314-398X]{Jeremy L. Smallwood}
\affiliation{ Department of Physics \& Astronomy, Baylor University, Waco, Texas 76798-7316, USA}
\affiliation{ Center for Astrophysics, Space Physics, and Engineering Research, Baylor University, Waco, Texas 76798-7316, USA}
\email{drjeremysmallwood@gmail.com}

\author[0000-0002-9442-137X]{Shang-Fei Liu (刘尚飞)}
\affiliation{School of Physics and Astronomy, Sun Yat-sen University, Zhuhai 519082, People's Republic of China}
\affiliation{CSST Science Center for the Guangdong-Hong Kong-Macau Greater Bay Area, Sun Yat-sen University, Zhuhai 519082, People's Republic of China}
\email[show]{liushangfei@mail.sysu.edu.cn}

\author[0000-0002-8400-0969]{Alessia Franchini}
\affiliation{Dipartimento di Fisica ``G. Occhialini'', Universit\`a degli Studi di Milano-Bicocca, Piazza della Scienza 3, 20126 Milano, Italy}
\email{alessia.franchini@unimib.it}

\correspondingauthor{Ruiqi Yang}
\begin{abstract}

We present three-dimensional smoothed particle hydrodynamics simulations of highly misaligned circumbinary disks (CBDs) around moderately eccentric equal-mass binaries ($e_\mathrm{b}=0.5$). We show that the binary accretion is modulated on the binary orbital period and exhibits two pulses near periastron. The dominant pulse peaks before periastron for an initial binary-disk misalignment of $60^\circ$, shifts to after periastron at $90^\circ$, and occurs at an even later post-periastron phase at $120^\circ$. We further show that the two pulses are accompanied by a time-dependent response of the circumstellar disks (CSDs) and by different distributions of accreting material within the cavity and around the CSDs. The qualitative pre- versus post-periastron distinction is also present in individual binary orbits despite variations in pulse amplitude. Our results motivate future tests of whether pulse timing is related to binary-disk orientation.
\end{abstract}



\section{Introduction} \label{sec:intro}
Circumbinary disks (CBDs) play an important role across a wide range of astrophysical environments, from protoplanetary disks around young stellar binaries to accretion disks around massive binary black holes \citep{barnes1998,kley2012}. Because many stars form in clusters and a large fraction belong to binary or higher-order multiple systems, both circumbinary and circumstellar disks (CSDs) are expected to be common \citep{lada2003,duchene2013}. Interactions among the binary, the CBD, and the CSDs can influence accretion onto the binary, orbital evolution, and the formation of circumbinary planets (see \citealt{lai2023a} for a review). Observations have identified CBDs in young systems such as DQ Tau, AK Sco, UZ Tau E, and [BHB2007] 11 \cite[e.g.][]{mathieu1997, czekala2016, alencar2003, anthonioz2015, czekala2015, simon2000a, prato2002, jensen2007,alves2019a}.

The gravitational torque exerted by a binary carves a central cavity in the CBD and regulates gas inflow through narrow streams. Early analytic and numerical work showed that gas could penetrate this cavity and feed the CSDs in a time-dependent manner \cite[e.g.,][]{artymowicz1994a,artymowicz1996}. Observationally, pre-main-sequence binaries such as TWA 3A exhibit periodic pulsed accretion \citep{tofflemire2017b,tofflemire2019}: the accretion rate rises by factors of several near periastron, broadly supporting the theoretical picture of periastron-driven pulsed accretion. Numerical studies further show that, for coplanar CBDs around circular binaries, a density lump often forms near the cavity edge, leading to accretion variability with a dominant period of about five binary orbits \citep{macfadyen2008,farris2014a,shi2012,munoz2016a,miranda2017,moody2019,munoz2020a,franchini2023,cocchiararo2024,franchini2024}. By contrast, for eccentric binaries, the dominant variability typically occurs on the binary orbital period \citep{munoz2016a}, with the accretion pattern and orbital evolution depending sensitively on binary eccentricity and mass ratio \citep{zrake2021a,siwek2023}.

However, CBDs are not necessarily coplanar. Misaligned CBDs may arise through turbulent fragmentation or chaotic accretion of misaligned material, and they have been observed in several systems \citep{bate2003, bate2018, czekala2019,ceppi2024}. For example, the pre-main-sequence binary KH 15D hosts a CBD tilted by $3^\circ$-$15^\circ$ \citep{chiang2004,winn2004,capelo2012,smallwood2019,poon2021}; IRS 43 contains a CBD misaligned by about $60^\circ$ \citep{brinch2016a}; 99 Herculis has a nearly polar debris CBD \citep{kennedy2012, smallwood2020}, and HD 98800 B hosts a nearly polar gaseous CBD \citep{kennedy2019a}. Theory predicts that misaligned CBDs around eccentric binaries undergo nodal precession about the binary eccentricity vector, and viscous dissipation drives them toward either coplanar or polar alignment \citep{verrier2009,farago2010, doolin2011, martin2017, lubow2018, zanazzi2018,smallwood2019}.

Despite the growing number of known misaligned CBDs, short-term accretion in such systems remains relatively unexplored. \citet{hayasaki2013} carried out three-dimensional smoothed particle hydrodynamics (SPH) simulations of accretion onto binary black holes from misaligned CBDs. They considered initial binary-disk tilts of $30^\circ$ and $45^\circ$ and found that, for circular binaries, the accretion rate exhibits two peaks per orbit because each black hole crosses the disk plane twice, whereas for eccentric binaries $(e>0.3)$, the eccentricity dominates and the accretion rate becomes single-peaked. However, their simulations employed only $\approx 5-9\times10^4$ SPH particles and did not explore higher misalignments approaching polar configurations.

In this paper, we present high-resolution SPH simulations of eccentric equal-mass binaries ($e_\mathrm{b}=0.5$) surrounded by misaligned CBDs with initial binary-disk tilts of $60^{\circ}$, $90^{\circ}$, and $120^{\circ}$. We find double-pulsed accretion near periastron in all three cases. The dominant pulse shifts to later orbital phases as the initial tilt increases. Our simulations also indicate that the CSDs can be periodically distorted near periastron and that the two pulses are accompanied by changes in the spatial distribution of accreting material within the cavity and around the CSDs. These results suggest that highly misaligned CBDs can produce accretion variability qualitatively different from that found in the mildly tilted and coplanar cases studied previously.

This paper is organized as follows. In Section \ref{sec:setup}, we describe the numerical method and simulation setup. In Section~\ref{sec:results}, we present the main results on short-term accretion variability and the associated accretion flow pattern. In Section \ref{sec:dis}, we discuss the observational implications, the orbital response of the CSDs, and the limitations of the models. In Section \ref{sec:con}, we summarize our conclusions.

\section{Hydrodynamical Simulation Setup} \label{sec:setup}
We perform numerical simulations using the SPH code {\sc phantom} \citep{price2018}. The {\sc phantom} code has been widely used to study misaligned CBDs \cite[e.g.,][]{nixon2015,martin2017,franchini2019,abod2022,smallwood2021a,smallwood2022,smallwood2025}.

We model an equal-mass binary with eccentricity $e_\mathrm{b}=0.5$. The binary is modeled with two sink particles \citep{bate1995} with total mass $M_\mathrm{b} = M_1 + M_2$, where $M_1$ and $M_2$ are the primary and secondary masses, respectively. We denote the binary semi-major axis as $a_\mathrm{b}$. Each sink has an accretion radius $r_\mathrm{acc}=0.05a_\mathrm{b}$. Gas particles that satisfy the accretion checks are accreted onto the sink, with their mass and angular momentum added accordingly \citep{price2018}. Any particle that enters within $0.8r_\mathrm{acc}=0.04a_\mathrm{b}$ is accreted immediately. Particles between $0.8r_\mathrm{acc}$ and $r_\mathrm{acc}$ are instead accreted if their specific angular momentum is smaller than that of a Keplerian orbit at the accretion radius and they are gravitationally bound to the sink, provided that they are more strongly bound to this sink than to any other. Initially, the binary lies in the $x$-$y$ plane and starts at apastron, with the binary eccentricity vector aligned with the $x$-axis.

The CBD is initially tilted with respect to the binary orbital plane by an angle $i_0$, where we consider $i_0\in\{60^\circ,\,90^\circ,\,120^\circ\}$. The disk is modeled with $2\times10^6$ equal-mass SPH particles with a total mass $M_\mathrm{d}=10^{-3}M_\mathrm{b}$. Its inner and outer radii are $r_\mathrm{in}=1.6a_\mathrm{b}$ and $r_\mathrm{out}=4.6a_\mathrm{b}$, respectively. Disks evolving towards polar alignment precess about the binary eccentricity vector rather than the binary angular momentum vector. Their longitude of ascending node therefore oscillates around $90^\circ$. We thus set the initial longitude of ascending node to $90^\circ$. The adopted inner radius approximately corresponds to the tidal truncation radius for polar disks \citep{miranda2015, franchini2019} and is smaller than that for coplanar disks \citep{artymowicz1994a}. This choice allows gas near the CBD inner edge to drift inward viscously and supply the CSDs \citep{smallwood2021a}.

We adopt initial conditions similar to those of \citet{smallwood2021a, smallwood2023a} and \citet{yang2026}. The initial surface density profile is $\Sigma(r)=\Sigma_0\left(r/r_\mathrm{in}\right)^{-3/2}$, where $\Sigma_0=1.13\times10^{-4}M_\mathrm{b}/a_\mathrm{b}^2$ is the initial surface density at the disk inner radius, and $r$ is the spherical radius. The disk pressure scale height is $H = c_\mathrm{s}/\Omega\propto r^{3/4}$, where $\Omega = (GM_\mathrm{b}/r^3)^{1/2}$. The disk aspect ratio at $r_\mathrm{in}$ is $H/r=0.1$. The kinematic viscosity is prescribed using the \citet{shakura1973} viscosity $\alpha_\mathrm{SS}$, such that $\nu=\alpha_\mathrm{SS}c_\mathrm{s} H$. We adopt $\alpha_\mathrm{SS}=0.1$ to enhance the inflow rate from the CBD into the binary region and thereby improve the resolution of the CSDs \cite[e.g.,][]{smallwood2021a,smallwood2022}. The initial CBD is resolved with an average smoothing length per scale height $\langle h\rangle/H \approx 0.11$.

To further improve the resolution of the CSDs, we adopt the modified locally isothermal equation of state described by \citet{smallwood2021a}, based on the prescription of \citet{farris2014a}. The sound speed is given by 
\begin{equation}
    c_\mathrm{s}=\mathcal{F}c_\mathrm{s0}\left(\frac{a_\mathrm{b}}{M_1+M_2}\right)^{4/3}\left(\frac{M_1}{r_1}+\frac{M_2}{r_2}\right)^{4/3},
\end{equation}
where $r_1$ and $r_2$ are the distances to the primary and secondary stars, respectively, and $c_\mathrm{s0}$ is a constant with dimensions of velocity. Here $\mathcal{F}$ is a dimensionless function of radius defined by
\begin{equation}
\mathcal{F}=\begin{cases}
\sqrt{0.001}, & \text{if }\ r_1\ \text{or}\ r_2\le r_\mathrm{c},\\
1, & \text{otherwise},
\end{cases}
\end{equation}
where $r_\mathrm{c}$ is the cutoff radius. The role of $\mathcal{F}$ is to reduce the sound speed around each binary component, thereby lengthening the viscous timescale and increasing both the mass and the resolution of the CSDs. Based on the results from \citet{smallwood2021a,smallwood2023a} and the analytical predictions of \citet{miranda2015}, the outer radius of the CSDs is around $0.35a_\mathrm{b}$, comparable to the Roche-lobe radius of an equal-mass binary evaluated at a separation of $a_\mathrm{b}$. We therefore set $r_\mathrm{c}=0.35a_\mathrm{b}$. This modification increases the CSD masses and thereby improves their numerical resolution. With this prescription, the disk aspect ratio in the CSDs at $r = 0.1a_\mathrm{b}$ is $H/r\sim0.01$. 

For CSD analysis, we divide each CSD into 30 uniform bins in spherical radius measured from its host star and average over particles in the range $0.05a_\mathrm{b}\le r \le0.50a_\mathrm{b}$, since the CSDs are truncated within this radius \citep{miranda2015}. For CBD analysis, we divide the disk into 300 uniform bins in spherical radius measured from the binary barycenter and average over particles in the range $1.5a_\mathrm{b}\le r\le10a_\mathrm{b}$. Within each bin, we calculate azimuthally averaged quantities such as surface density, tilt, longitude of the ascending node, and eccentricity.

To examine the instantaneous geometry of the CSDs and CBDs, we calculate their mass-weighted angular momentum and coherent eccentricity vectors. The coherent eccentricity magnitude $e_D$, defined in Appendix~\ref{app:geometry_robustness}, is used to assess whether the particle eccentricity vectors define a coherent apsidal direction; it is distinct from the density-weighted scalar CSD eccentricity $e$ shown in Figure~\ref{fig:inc_ecc_mass}. For the CSD vector analysis introduced here, $k=1$ and $2$ denote the circumprimary and circumsecondary disks, respectively, and $r_k$ is the distance from a gas particle to star $k$. We select particles with $0.05a_\mathrm{b}\leq r_k\leq0.35a_\mathrm{b}$ that are bound to the host star and more strongly bound to it than to the companion. For the CBD, we analyze a cavity-edge annulus, $1.5a_\mathrm{b}\leq r<2.5a_\mathrm{b}$, and an outer annulus, $2.5a_\mathrm{b}\leq r<4.0a_\mathrm{b}$.

We measure the CSD orientation only when at least 40 particles satisfy the selection criteria, to avoid poorly sampled estimates. For CSD $k$, we measure the apsidal orientation only when the magnitude of the coherent eccentricity vector satisfies $e_k\geq0.02$, so that the apsidal direction is sufficiently well defined. The vector and angular definitions are given in Appendix~\ref{app:geometry_robustness}.

The fiducial simulations are evolved to $25\,P_\mathrm{b}$. For the analysis of short-term accretion variability, we focus on the interval $10$-$20\,P_\mathrm{b}$. In all three cases, the accretion rate rises during the first several binary orbits as gas from the initial CBD spreads inward and begins to feed the binary region. Excluding the first $10\,P_\mathrm{b}$ therefore reduces the influence of the initial ramp-up to the phase-folded curves.

The upper limit of $20\,P_\mathrm{b}$ is chosen because the $i_0=60^\circ$ and $120^\circ$ disks evolve rapidly toward the polar configuration. Since our goal is to compare the accretion response for highly misaligned prograde, polar, and highly misaligned retrograde disk orientations, averaging over a longer interval would mix the intended tilt dependence with secular disk reorientation. The $10$-$20\,P_\mathrm{b}$ interval therefore provides a controlled short-term comparison after the initial accretion ramp-up, while the three disks still retain their distinct prograde, polar, and retrograde orientations.

\section{Results}
\label{sec:results}
\subsection{Short-Term Accretion Variability}

\begin{figure}
    \centering
    \includegraphics[width=\linewidth]{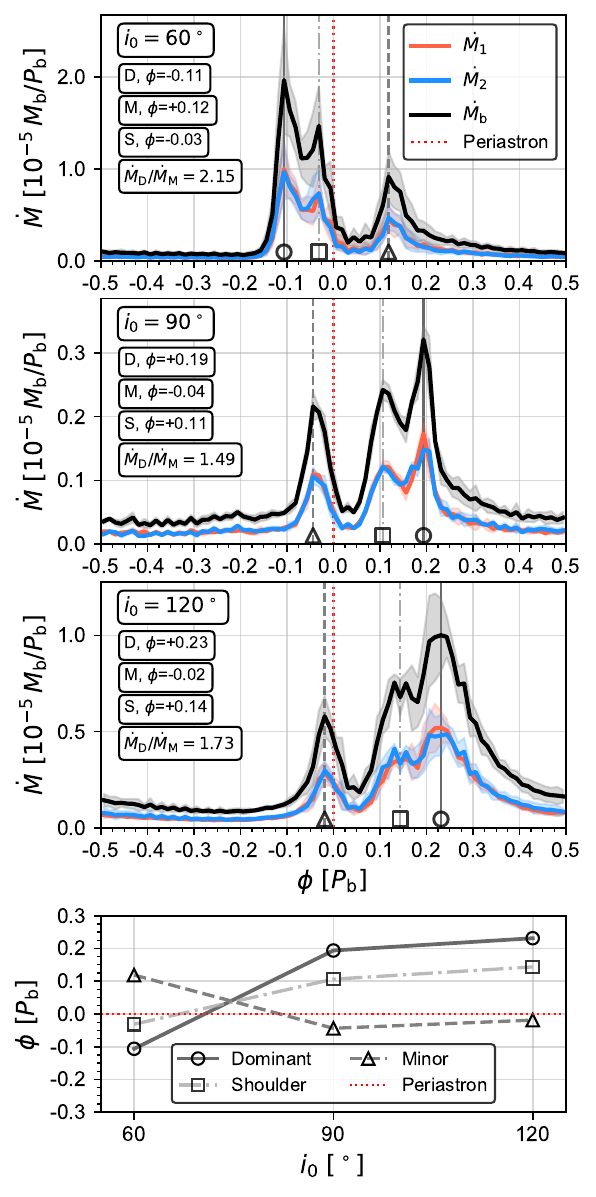}
    \caption{Accretion rates onto the primary ($\dot{M}_1$, red), secondary ($\dot{M}_2$, blue), and binary ($\dot{M}_\mathrm{b}$, black) as a function of orbital phase ($\phi$) for three initial CBD tilts. The curves are averaged over $10$-$20\,P_\mathrm{b}$, and the shaded regions show the $1\sigma$ dispersion. The dotted red lines mark periastron $\phi=0\,P_\mathrm{b}$. The dominant (D; circle), shoulder (S; square), and minor (M; triangle) accretion features are identified from a weakly smoothed phase-averaged binary accretion curve using a three-bin Savitzky--Golay filter ($\Delta\phi=0.0375\,P_\mathrm{b}$). The smoothing is used only to determine the pulse phases. The lower panel summarizes the pulse phases as a function of initial tilt. All three cases exhibit double-pulsed accretion near periastron. The dominant peak shifts to later phases as the initial tilt increases.}
    \label{fig:acc_phase}
\end{figure}

\begin{figure}
    \centering
    \includegraphics[width=\linewidth]{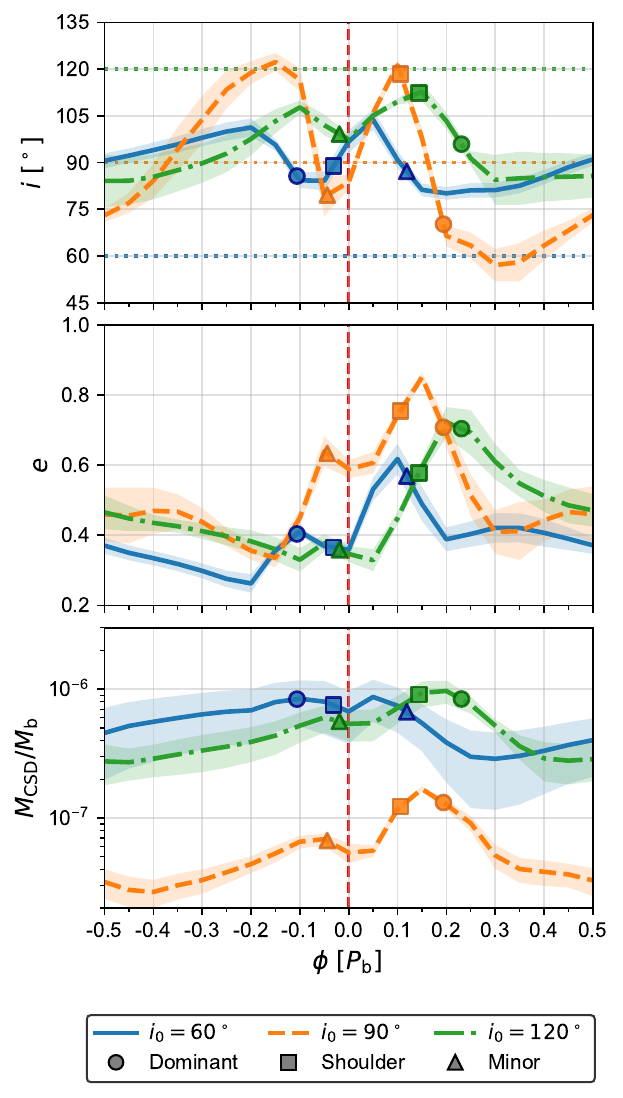}
    \caption{Orbital-phase evolution of the density-weighted circumstellar disk tilt $i$ (top), eccentricity $e$ (middle), and mass $M_\mathrm{CSD}/M_\mathrm{b}$ (bottom). The curves are averaged over $10$-$20\,P_\mathrm{b}$, and the shaded regions show the $1\sigma$ dispersion. The blue solid, orange dashed, and green dash-dotted curves correspond to initial CBD tilts of $60^\circ$, $90^\circ$, and $120^\circ$, respectively. The vertical dashed red line marks periastron. In the top panel, the horizontal dotted lines mark the corresponding initial CBD tilts. Circles, squares, and triangles indicate the dominant, shoulder, and minor accretion-pulse phases identified from Figure~\ref{fig:acc_phase}, with colors matched to the corresponding initial tilt. These CSD diagnostics show strong orbital modulation, indicating a time-dependent CSD response on the binary orbital period.}
    \label{fig:inc_ecc_mass}
\end{figure}

Figure~\ref{fig:acc_phase} shows the accretion rates onto the individual stars and the binary as a function of orbital phase for disks with initial tilts of $60^\circ$, $90^\circ$, and $120^\circ$. The curves are averaged over $10$-$20\,P_\mathrm{b}$, and the shaded regions indicate the 1$\sigma$ dispersion. The red, blue, and black curves denote the accretion rates onto the primary, the secondary, and the binary, respectively. The dotted red lines mark periastron, with $\phi=0\,P_\mathrm{b}$ defined as the periastron passage. The figure also marks the dominant, shoulder, and minor pulses identified from the binary accretion curve, together with the ratio between the dominant and minor pulses. The pulse phases are identified from a weakly smoothed version of the phase-averaged accretion curve. We fold the accretion rate around periastron and average it into 80 phase bins over $-0.5\,P_\mathrm{b}\le\phi\le0.5\,P_\mathrm{b}$. For phase identification only, we smooth the binary accretion curve with a Savitzky--Golay filter using a three-bin window ($\Delta\phi=0.0375\,P_\mathrm{b}$) and a second-order polynomial. The dominant and minor pulses are defined as the two strongest local maxima on opposite sides of periastron, while the shoulder is defined as the enhancement located between the dominant pulse and periastron. For the $i_0=120^\circ$ case, where the shoulder contains multiple nearby sub-peaks, we use their amplitude-weighted mean phase. The smoothing is used only to identify the pulse phases, and the plotted accretion rates and peak ratios are measured from the unsmoothed phase-averaged curves.

In all three cases, the binary accretion rate exhibits two pulses near periastron. For $i_0=60^\circ$, the dominant pulse occurs before periastron (\textit{pre-periastron}), peaking at phase $\phi\approx-0.11\,P_\mathrm{b}$, followed by a shoulder near $\phi\approx-0.03\,P_\mathrm{b}$. A weaker post-periastron pulse occurs at $\phi\approx+0.12\,P_\mathrm{b}$. For $i_0=90^\circ$, the dominant pulse shifts to after periastron (\textit{post-periastron}) and peaks at $\phi\approx+0.19\,P_\mathrm{b}$, while the weaker pre-periastron pulse is located at $\phi\approx-0.04\,P_\mathrm{b}$, and the shoulder appears at $\phi\approx+0.11\,P_\mathrm{b}$. For $i_0=120^\circ$, the dominant post-periastron pulse is delayed further, peaking at $\phi\approx+0.23\,P_\mathrm{b}$, with a shoulder at $\phi\approx+0.14\,P_\mathrm{b}$ and a weak minor pulse at $\phi\approx-0.02\,P_\mathrm{b}$. The lower panel of Figure~\ref{fig:acc_phase} shows that the dominant pulse shifts to later phases as the initial tilt increases.

The relative strength of the two pulses also differs among the three simulations. The dominant-to-minor peak ratio is largest for $i_0=60^\circ$ ($\dot{M}_\mathrm{D}/\dot{M}_\mathrm{M}\approx2.15$), smaller for $i_0=90^\circ$ ($\dot{M}_\mathrm{D}/\dot{M}_\mathrm{M}\approx1.49$), and intermediate for $i_0=120^\circ$ ($\dot{M}_\mathrm{D}/\dot{M}_\mathrm{M}\approx1.73$). The $i_0=60^\circ$ and $120^\circ$ cases also show larger dispersion than the $90^\circ$ case, consistent with their stronger secular tilt oscillation. In addition, the mean accretion rates in the $i_0=60^\circ$ and $120^\circ$ cases are higher than in the $90^\circ$ case. This trend is consistent with previous studies suggesting that disks misaligned away from the polar alignment generally accrete more efficiently onto the binary \citep{nixon2013, smallwood2022,smallwood2025,yang2026}.

Figure~\ref{fig:inc_ecc_mass} shows the phase-averaged evolution of the density-weighted circumstellar disk tilt $i$, eccentricity $e$, and mass $M_\mathrm{CSD}/M_\mathrm{b}$ over $10$-$20\,P_\mathrm{b}$. The circumprimary and circumsecondary disks behave similarly, so we present only the circumprimary disk. The blue solid, orange dashed, and green dash-dotted curves correspond to initial CBD tilts of $60^\circ$, $90^\circ$, and $120^\circ$, respectively. The shaded regions show the $1\sigma$ dispersion, while the markers identify the dominant, shoulder, and minor accretion-pulse phases from Figure~\ref{fig:acc_phase}.

The phase-averaged CSD tilt can vary substantially with orbital phase. For $60^\circ$, the CSD tilt ranges from about $80^\circ$ to $105^\circ$, exceeding the corresponding CBD tilt of $60^\circ$. For $90^\circ$, the tilt varies over a wider range, from about $50^\circ$ to $125^\circ$. For $120^\circ$, the tilt varies moderately, between $75^\circ$ and $115^\circ$. The colored horizontal dashed lines in the top panel emphasize that the CSD tilt is generally offset from the initial CBD tilt and varies substantially with orbital phase. The especially large tilt variation in the $90^\circ$ case may partly reflect a greater depletion of CSD material. By contrast, the $60^\circ$ and $120^\circ$ cases show more persistent CSD structures, consistent with their more moderate tilt variations. Notably, the identified accretion pulses occur near phases where the CSD tilt changes rapidly, suggesting that accretion pulses are accompanied by a strongly time-dependent CSD orientation.

The CSD eccentricity is also modulated on the binary orbital period. The eccentricity tends to increase around periastron and near the phases of the identified accretion features, although the dominant accretion pulse does not always coincide exactly with the largest eccentricity. For $i_0=60^\circ$, the dominant pulse occurs before periastron, when the eccentricity is enhanced. For $i_0=90^\circ$ and $120^\circ$, the dominant pulse occurs after periastron and is associated with the eccentricity enhancement.

The CSD mass shows a similarly strong orbital modulation. For $60^\circ$, $M_\mathrm{CSD}$ remains relatively high before periastron, close to the phase of the dominant pre-periastron pulse. For $90^\circ$, the CSD mass is substantially lower overall, but still shows a modest enhancement around the dominant post-periastron pulse. For $120^\circ$, the CSD mass rises after periastron and reaches its maximum near the dominant pulse. Taken together, Figure~\ref{fig:inc_ecc_mass} shows that the CSD tilt, eccentricity, and mass are all modulated on the binary orbital period, indicating that the CSDs are repeatedly distorted by the interaction with the companion on an eccentric orbit. These variations occur near the accretion features identified in Figure~\ref{fig:acc_phase}, suggesting that the short-term accretion variability is accompanied by a strongly time-dependent CSD response.

We performed additional tests to assess whether averaging over $10$-$20\,P_\mathrm{b}$ masks variations among individual binary orbits. Appendix~\ref{app:geometry_robustness} shows that the qualitative pre- versus post-periastron distinction is preserved in the individual orbits, so the adopted averaging interval does not qualitatively alter the main accretion trend reported above.

\begin{figure*}
    \centering
    \begin{interactive}{animation}{e0p5i60_animation.mp4}
        \includegraphics[width=0.95\linewidth]
        {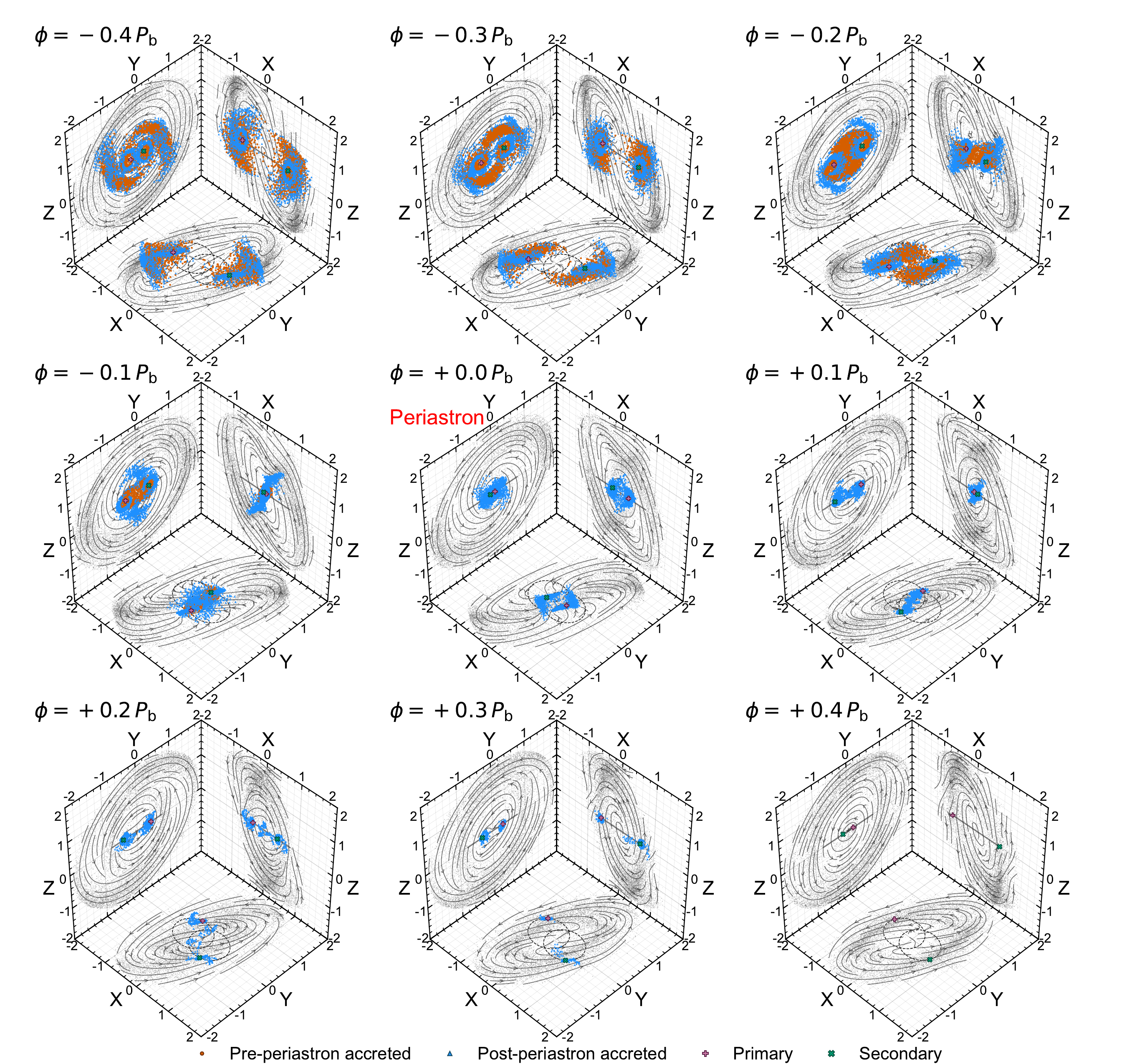}
    \end{interactive}
    \caption{Gas inflow morphology for the $i_0=60^\circ$ case over one representative orbit ($15$-$16\,P_\mathrm{b}$). Each panel shows one snapshot, spaced by $0.1\,P_\mathrm{b}$, from $\phi=-0.4\,P_\mathrm{b}$ to $\phi=+0.4\,P_\mathrm{b}$. Three orthogonal projections of the inner disk region are shown in each snapshot. A representative subset of particles accreted before (after) periastron is shown as orange circles (blue triangles), while a subsample of the remaining gas particles is shown in gray. The primary and secondary sinks are marked by magenta plus and green X symbols, respectively, and are plotted above the gas particles. Dashed black curves indicate the projected binary orbits. Dark-gray streamlines show the projected gas-velocity field, computed from all particles within the plotted region. In this case, the dominant pre-periastron pulse appears to be associated with CSD-mediated capture, as the two CSDs become strongly distorted and develop a transient bridge structure near periastron. An animated version of this figure is available in the HTML article. The animation contains 50 snapshots spanning $-0.5\leq\phi/P_\mathrm{b}<+0.5$, and shows the continuous evolution of the tracked particles, binary components and projected orbits, and projected gas-velocity field.}
    \label{fig:i60_flow}
\end{figure*}

\begin{figure*}
    \centering
    \begin{interactive}{animation}{e0p5i90_animation.mp4}
        \includegraphics[width=0.95\linewidth]
        {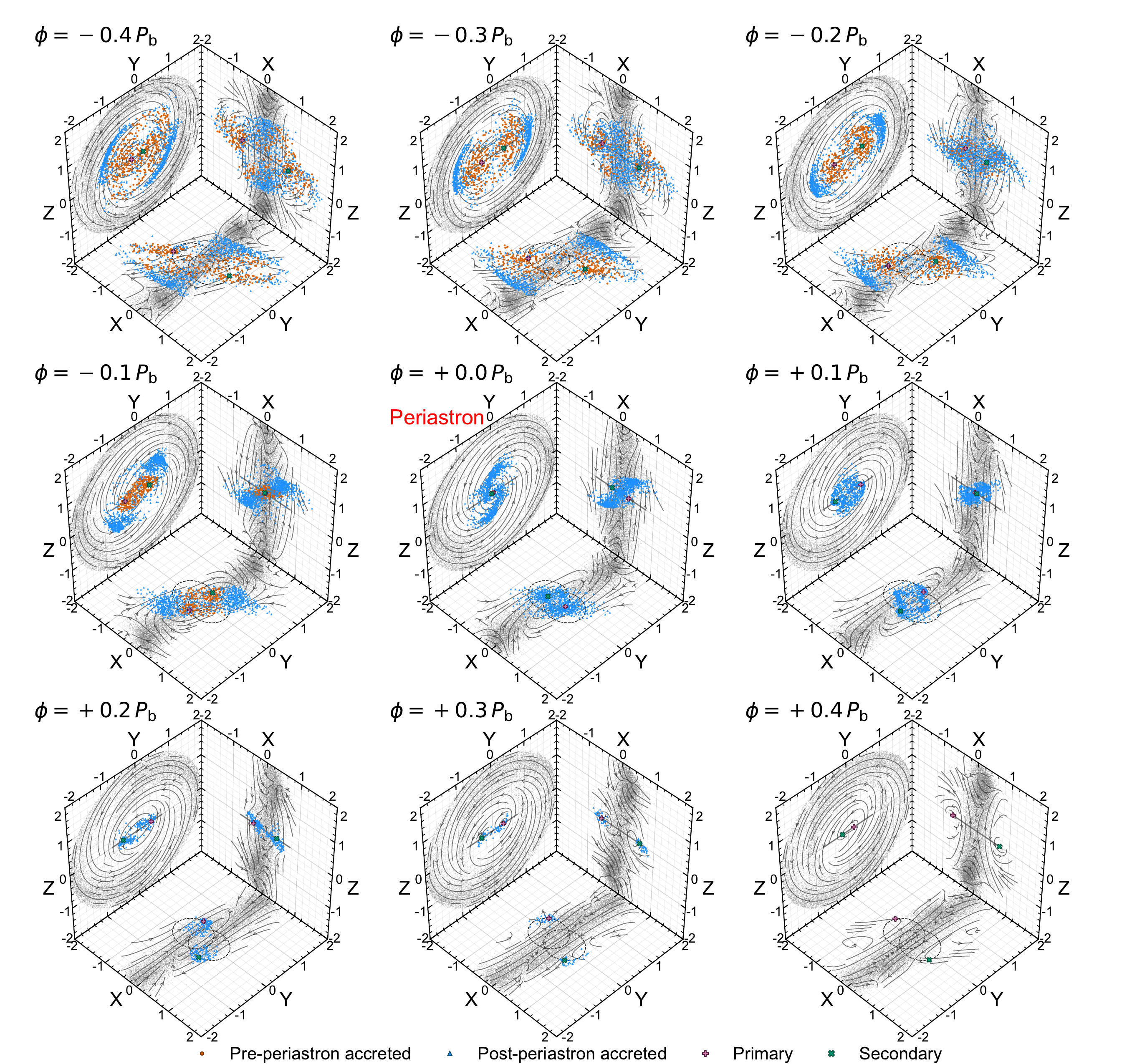}
    \end{interactive}
    \caption{Same as Figure~\ref{fig:i60_flow}, but for $i_0=90^\circ$. The pre-periastron component is associated with gas displaced from the CBD inner edge and subsequently falling back toward the binary, whereas the stronger post-periastron pulse is accompanied by a larger concentration of accreting material near the cavity center and around the stars. An animated version of this figure is available in the HTML article and follows the same format, phase range, and snapshot sampling as the animation for Figure~\ref{fig:i60_flow}}.
    \label{fig:i90_flow}
\end{figure*}

\begin{figure*}
    \centering
    \begin{interactive}{animation}{e0p5i120_animation.mp4}
        \includegraphics[width=0.95\linewidth]
        {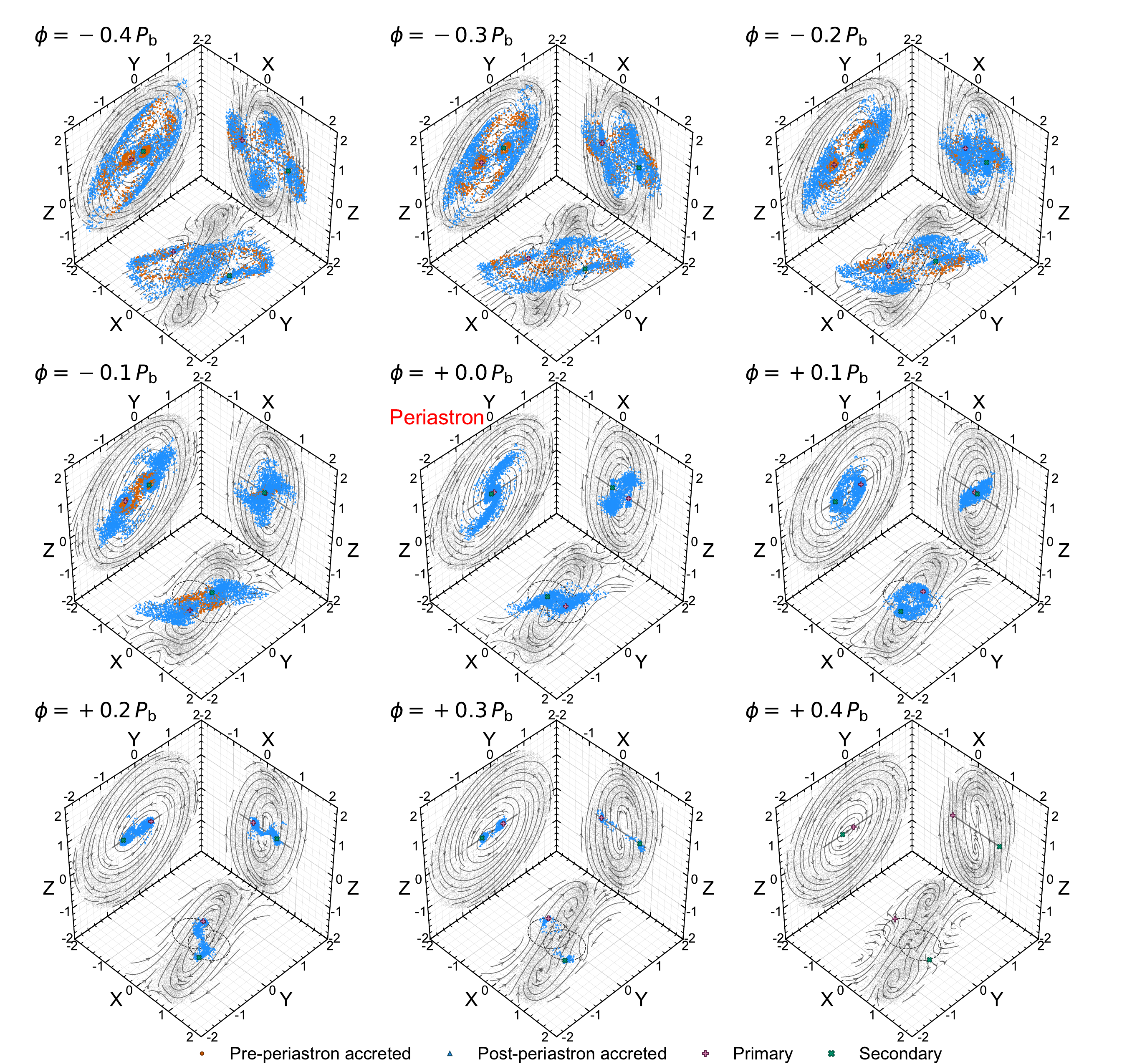}
    \end{interactive}
    \caption{Same as Figure~\ref{fig:i60_flow}, but for $i_0=120^\circ$. The pre-periastron pulse is weak, whereas the dominant post-periastron pulse is associated with a larger amount of accreting material retained in the cavity and near the CSD region. An animated version of this figure is available in the HTML article and follows the same format, phase range, and snapshot sampling as the animation for Figure~\ref{fig:i60_flow}}.
    \label{fig:i120_flow}
\end{figure*}

\subsection{Accretion Flows from CBD onto the Binary}

To connect the phase-dependent accretion variability with the gas dynamics, we examine one representative binary orbit and tag particles accreted during that orbit ($15$-$16\,P_\mathrm{b}$). Figures~\ref{fig:i60_flow}, \ref{fig:i90_flow}, and \ref{fig:i120_flow} show three orthogonal projections of the inner disk region at nine phases spaced by $0.1\,P_\mathrm{b}$ from $\phi=-0.4\,P_\mathrm{b}$ to $\phi=+0.4\,P_\mathrm{b}$. A representative subset of particles accreted before (after) periastron is plotted as orange circles (blue triangles), while a subsample of the remaining gas is shown in gray. The primary and secondary sinks are marked by magenta plus and green X symbols, respectively, and are plotted above the gas particles to clarify the binary position and orbital phase. The projected binary orbits are indicated by dashed black curves. Dark-gray streamlines indicate the projected gas velocity field, which is computed from all gas particles within the plotted inner region. The corresponding animations extend the sampling to 50 snapshots spanning the complete orbit.

For $i_0=60^\circ$, the pre-periastron accretion appears to be dominated by material already bound to, or rapidly captured by the CSDs. From $\phi=-0.4\,P_\mathrm{b}$ to $\phi=-0.2\,P_\mathrm{b}$, the pre-periastron-tagged particles progressively concentrate around the CSDs rather than remaining distributed along the CBD inner edge. As the binary approaches periastron, the two CSDs become increasingly distorted and develop a transient bridge. Most of these particles are accreted during this phase, consistent with the dominant pulse being associated with CSD-mediated capture and direct CSD interaction. After periastron, the post-periastron-tagged particles are less abundant and remain closer to the stars, consistent with a weaker minor pulse. From $\phi=+0.2\,P_\mathrm{b}$ to $\phi=+0.4\,P_\mathrm{b}$, the streams from the inner CBD weaken and the bridge dissipates.

For $i_0=90^\circ$, the picture differs qualitatively. Before periastron, pre-periastron-tagged particles appear to come from gas displaced from the CBD inner edge and launched out of the CBD plane. The displaced material subsequently falls back toward the binary, consistent with the fallback picture described by \citet{smallwood2023a}. In contrast, after periastron the post-periastron-tagged particles become more prominent near the cavity center and around the stars, matching the stronger post-periastron pulse seen in Figure~\ref{fig:acc_phase}. In this case, the pre- and post-periastron pulses are associated with different spatial distributions of accreting material, with the latter showing a larger concentration of material near the cavity center and around the stars.

For $i_0=120^\circ$, the asymmetry between the dominant and minor pulses is most pronounced. The pre-periastron-tagged particles are fewer in number and more localized near the binary than in the lower-tilt cases, consistent with the weak pre-periastron pulse. By contrast, post-periastron-tagged particles are more abundant and show a clearer connection to gas at the CBD inner edge, especially from periastron to $\phi\approx+0.20\,P_\mathrm{b}$. The dominant post-periastron pulse therefore appears to be associated with a larger amount of material retained in the cavity and near the CSD region.

Taken together, these snapshots show that the three models differ in the relative strength and timing of the two pulses, as well as the spatial distribution of the accreting material. The $i_0=60^\circ$ case shows a stronger association between the dominant pre-periastron pulse and material concentrated near the CSDs, together with transient CSD distortion around periastron. In contrast, the $i_0=90^\circ$ and $120^\circ$ cases both show stronger post-periastron accretion from material located in the cavity and near the CSD region. Compared with the $i_0=90^\circ$ case, the larger post-periastron pulse in the $i_0=120^\circ$ case appears to be associated with a greater concentration of accreting material retained in the cavity and near the CSD region.

\section{Discussion} \label{sec:dis}
\subsection{Implications for observations}

Our simulations suggest that highly misaligned CBDs around eccentric binaries can produce double-pulsed accretion near periastron. In our models, both the phase and relative strength of the two pulses vary with the CBD tilt (Figure~\ref{fig:acc_phase}). The dominant pulse occurs before periastron for $i_0=60^\circ$ and after periastron for $i_0=90^\circ$ and $i_0=120^\circ$, with the dominant pulse in the $i_0=120^\circ$ model occurring later than that in the $i_0=90^\circ$ case. These differences motivate future tests of whether pulse timing is related to binary-disk orientation, but do not yet define a calibrated observational diagnostic.

Highly misaligned CBDs can undergo secular tilt oscillations and reorientation \citep{verrier2009,farago2010,doolin2011,martin2017,lubow2018,zanazzi2018,smallwood2019}. The inclinations explored here may therefore resemble configurations encountered at different stages of a disk's secular evolution. However, our three simulations were initialized independently and should not be interpreted as successive snapshots along a single demonstrated evolutionary track. They instead provide a controlled comparison of the short-term accretion response at three different disk orientations.

An observational test of this possible trend would require an accurate binary orbital solution, together with an independent constraint on the disk geometry. Radial-velocity monitoring, eclipses or transits, and astrometry can constrain the binary eccentricity, orbital orientation, and time of periastron passage \citep[e.g.,][]{tofflemire2017b,tofflemire2019,czekala2021a}. These measurements provide the phase reference needed to phase-fold accretion measurements and determine the timing of an accretion pulse relative to periastron rather than to an arbitrary observational phase. Repeated observations over several binary orbits would also be important for distinguishing a persistent phase dependence from orbit-to-orbit variability.

Time-domain spectroscopy and photometric monitoring provide practical ways to measure this short-term variability. High-resolution spectroscopy with facilities such as VLT/X-shooter \citep{vernet2011,rigliaco2012} can measure the orbital-phase dependence of H$\alpha$ and other accretion tracers, while HST/COS \citep{green2011} can probe ultraviolet accretion diagnostics. Photometric monitoring can also identify the timing and duration of accretion pulses. For example, \citet{tofflemire2017b,tofflemire2019} monitored the classical T Tauri binary TWA 3A and found accretion flares peaking near periastron, with enhancements ranging from a factor of a few to approximately an order of magnitude. TWA 3A is inferred to host a nearly coplanar CBD \citep{czekala2021a} and is therefore not represented by our highly misaligned models. We do not apply our simulated pulse phases to TWA 3A or use them to infer its disk orientation. Rather, TWA 3A demonstrates that orbital-phase-dependent accretion can be measured in an eccentric young binary. Similar monitoring of future eccentric systems with independently established disk misalignments could test whether similar pulse-timing differences occur in real systems.

Spatially resolved imaging would provide complementary and independent constraints on the disk geometry. Polarimetric imaging with facilities such as SPHERE/ZIMPOL, together with ALMA continuum or molecular-line observations, can constrain the projected CBD orientation and cavity structure. The cavity size may provide additional information about binary-disk misalignment, since reduced tidal torques in misaligned disks can produce smaller inner cavities than in coplanar systems \citep{franchini2019}. However, the cavity structure also depends on other binary and disk properties and cannot by itself provide a unique measurement of the misalignment. Combined with phase-resolved accretion variability, such observations could test whether the dominant-pulse phase is associated with binary-disk orientation. A quantitative interpretation will require a broader simulation survey and system-specific comparisons.

\subsection{Orbital response of the circumstellar disks}

Our simulations show that the CSDs undergo strong variations in mass, eccentricity, and orientation on the binary orbital timescale. The acute angle between the two CSD planes generally remains small, indicating a largely coherent response of their orbital planes, while their apsidal orientations vary more strongly with orbital phase. The relevant CSD geometry is therefore three-dimensional and time-dependent and cannot be described as an interaction between two stationary disk planes. These variations are closely associated with the short-term accretion variability.

The structure and evolution of CSDs in eccentric binaries can depend on both the physical conditions and the numerical treatment. Coplanar simulations show that eccentric binaries can generate complex, phase-dependent cavity flows \citep{mosta2019}, while higher-resolution calculations demonstrate that CSD eccentricity and apsidal precession can depend on numerical resolution, computational-domain size, viscosity, and thermodynamics \citep{jordan2021}. Continued mass transfer from a CBD can also substantially affect the mass supply and long-term evolution of a CSD \citep{marzari2025}. At lower binary eccentricity, \citet{smallwood2021a,smallwood2023a} found that continuously fed, highly inclined CSDs can persist and undergo von Zeipel-Kozai-Lidov (ZKL, \citealt{zeipel1909,kozai1962,lidov1962}) cycles. Such cycles can produce exchanges between disk tilt and eccentricity and generate shocks \citep{fu2017a,smallwood2021a,smallwood2023a}. Simulations of polar CBDs indicate that CSD formation can also depend on sink radius and binary mass ratio \citep{chen2026}. Together, these studies emphasize that the CSD response is not determined by disk tilt or binary eccentricity alone.

The adopted values $H/r=0.1$ at the CBD inner edge and $\alpha_\mathrm{SS}=0.1$ describe a relatively thick and highly viscous CBD. The high viscosity was chosen to enhance gas delivery into the cavity and improve the numerical resolution of the CSDs. Our results therefore characterize the short-term CSD response in a strongly supplied and interacting regime. Determining the longer-term evolution of these CSDs and assessing implications for S-type planet formation will require simulations spanning a broader range of disk conditions and mass-supply rates.

\subsection{Limitations and future work}

Several limitations of the present results should be kept in mind. First, a broader exploration of parameter space is needed to determine where the dominant pulse switches from pre- to post-periastron, how the critical transition tilt depends on binary eccentricity and other parameters, and under what conditions double-pulsed accretion persists. The fiducial simulations are not evolved to a global viscous steady state, and the $10$-$20\,P_\mathrm{b}$ interval is therefore used to characterize the short-term orbital modulation rather than the converged long-term evolution of the CBD. Secular reorientation of the CBD \citep[e.g.,][]{martin2017,lubow2018,smallwood2019} can change the geometry of the CBD--binary--CSD interaction, so the pulse phases measured over $10$-$20\,P_\mathrm{b}$ should not be extrapolated to later stages.

Second, our simulations adopt a modified locally isothermal equation of state to increase the CSD mass and resolution. This prescription does not treat shock heating, radiative cooling, or thermal evolution self-consistently. Different cooling prescriptions can alter the circumbinary cavity structure and disk precession rate even in coplanar models \citep{sudarshan2022}. More realistic thermodynamics could also alter the CSD eccentricity, thickness, and shock structure \citep{li2021a,wang2023,wang2023b}. The robustness of both the CSD response and the accretion pulse to the thermodynamic treatment therefore remains to be tested.

Third, the absolute accretion rates, relative pulse amplitudes, and detailed CSD properties may depend on the sink prescription. Although our particle-based sink algorithm does not employ an independently specified sink-depletion rate or radial removal profile, the adopted accretion radius and particle-removal criteria still define an unresolved numerical boundary. \citet{dittmann2021} showed that the removal rate and radial profile of grid-based sinks can affect CSD structure and accretion variability. These parameters do not map directly onto our particle-based implementation. In our simulations, changing the accretion radius would be expected primarily to affect gas storage near the stars and the normalization of the measured accretion rate, although some dependence of the relative pulse amplitudes cannot be excluded. We therefore interpret the quantitative accretion amplitudes and detailed CSD properties in the context of the adopted value $r_\mathrm{acc}=0.05a_\mathrm{b}$.

Finally, synthetic observables, such as line profiles and light curves, would enable direct comparison with time-domain observations of young eccentric binaries and possibly with candidate supermassive black hole binaries.

\section{Conclusions} \label{sec:con}
In this paper, we have shown that an eccentric equal-mass binary in a highly misaligned circumbinary disk can exhibit pronounced orbit-modulated accretion variability with two pulses near periastron. The dominant pulse occurs before periastron in the $60^\circ$ case and after periastron in the $90^\circ$ and $120^\circ$ cases, with the $120^\circ$ pulse occurring later than the $90^\circ$ pulse.

We further show that the two pulses are accompanied by a time-dependent response of the CSDs and by different spatial distributions of accreting material within the cavity and around the CSDs. In the $60^\circ$ case, the dominant pre-periastron pulse is likely linked to material concentrated near the CSDs and to transient CSD distortion around periastron. In the $90^\circ$ and $120^\circ$ cases, the dominant pulse occurs after periastron and is associated with material retained in the cavity and near the CSD region, with the $120^\circ$ case showing a larger amount of post-periastron accreting material.

The qualitative pre- versus post-periastron distinction among the three models is also present in the individual binary orbits and is therefore not an artifact of phase averaging. The CSDs also exhibit pronounced orbital variations in mass, eccentricity, and orientation, while the acute angle between the two CSD planes generally remains small. These results motivate future tests of how pulse timing and longer-term CSD evolution depend on binary and disk properties.

\begin{acknowledgments}
We thank the anonymous reviewer for constructive comments and suggestions that have improved our work. We thank Douglas N. C. Lin for useful discussions. RY and S.-F.L. are supported by the National Natural Science Foundation of China (grant No. 42530203) and the Guangdong Basic and Applied Basic Research Foundation (grant No. 2021B1515020090).  We thank Daniel Price for providing the {\sc phantom} \citep{price2018} code for our SPH simulations. Numerical simulations were conducted on the Loong cluster at the School of Physics and Astronomy, Sun Yat-sen University, China.
\end{acknowledgments}

\software{Phantom \citep{price2018}, Matplotlib \citep{hunter2007}, NumPy \citep{harris2020}, Sarracen \citep{harris2023}, SciPy \citep{virtanen2020}, pandas \citep{mckinney2010}}

\appendix
\section{Disk Geometry and Robustness of Phase Averaging}
\label{app:geometry_robustness}

For a selected disk region $D$, we calculate the mass-weighted angular-momentum vector

\begin{equation}
\boldsymbol{L}_D
=
\sum_{j\in D}
m_j
\left(
\boldsymbol{r}_j
\times
\boldsymbol{v}_j
\right),
\qquad
\hat{\boldsymbol{\ell}}_D
=
\frac{\boldsymbol{L}_D}
{|\boldsymbol{L}_D|}.
\label{eq:disk_angmom}
\end{equation}

The coherent eccentricity vector is

\begin{equation}
\boldsymbol{e}_D
=
\frac{
\sum_{j\in D}
m_j\boldsymbol{e}_j
}{
\sum_{j\in D}m_j
},
\label{eq:disk_ecc}
\end{equation}

We denote the magnitude of this vector by $e_D\equiv|\boldsymbol{e}_D|$, which measures the degree to which the particle eccentricity vectors define a coherent apsidal direction. The osculating eccentricity vector of particle $j$ is

\begin{equation}
\boldsymbol{e}_j
=
\frac{
\boldsymbol{v}_j
\times
\left(
\boldsymbol{r}_j
\times
\boldsymbol{v}_j
\right)
}{
G M_\mathrm{c}
}
-
\frac{\boldsymbol{r}_j}
{|\boldsymbol{r}_j|}.
\label{eq:particle_ecc}
\end{equation}

For a CSD, positions and velocities are measured relative to the host star and $M_\mathrm{c}$ is the stellar mass. For the CBD, they are measured relative to the binary barycenter and $M_\mathrm{c}=M_\mathrm{b}$. Because the gas evolves in a pressure-supported, non-Keplerian, and time-dependent potential, we use $\boldsymbol{e}_D$ as a diagnostic of coherent apsidal orientation rather than as an exact Keplerian orbital element. For CSD $k$, we write $\hat{\boldsymbol{\ell}}_k$ and $\boldsymbol{e}_k$ for the corresponding $\hat{\boldsymbol{\ell}}_D$ and $\boldsymbol{e}_D$ quantities.

The relative orientation of the two CSD planes is quantified by the acute plane angle

\begin{equation}
i_{12}
=
\cos^{-1}
\left(
\left|
\hat{\boldsymbol{\ell}}_1
\boldsymbol{\cdot}
\hat{\boldsymbol{\ell}}_2
\right|
\right).
\end{equation}

The absolute value makes $i_{12}$ an angle between geometrical planes rather than between directed angular-momentum vectors.

Because a CSD is generally inclined relative to the binary orbital plane, we project both its coherent eccentricity vector and the periastron direction of the host star's barycentric orbit onto the instantaneous CSD plane. For any vector $\boldsymbol{a}$, we define

\begin{equation}
\mathcal{P}_k(\boldsymbol{a})
=
\boldsymbol{a}
-
\left(
\boldsymbol{a}
\boldsymbol{\cdot}
\hat{\boldsymbol{\ell}}_k
\right)
\hat{\boldsymbol{\ell}}_k.
\end{equation}

Let $\hat{\boldsymbol{q}}_k$ denote the unit vector toward periastron of the barycentric orbit of star $k$. The projected reference and eccentricity directions are

\begin{equation}
\hat{\boldsymbol{p}}_{\mathrm{ref},k}
=
\frac{
\mathcal{P}_k(\hat{\boldsymbol{q}}_k)
}{
\left|
\mathcal{P}_k(\hat{\boldsymbol{q}}_k)
\right|
},
\qquad
\hat{\boldsymbol{p}}_{e,k}
=
\frac{
\mathcal{P}_k(\boldsymbol{e}_k)
}{
\left|
\mathcal{P}_k(\boldsymbol{e}_k)
\right|
}.
\end{equation}

The signed apsidal angle $\Delta\varpi_k$ is the unique angle in $(-\pi,\pi]$ satisfying $\cos\Delta\varpi_k=\hat{\boldsymbol{p}}_{\mathrm{ref},k}\boldsymbol{\cdot}\hat{\boldsymbol{p}}_{e,k}$, and $\sin\Delta\varpi_k=\hat{\boldsymbol{\ell}}_k\boldsymbol{\cdot}\left(\hat{\boldsymbol{p}}_{\mathrm{ref},k}\times\hat{\boldsymbol{p}}_{e,k}\right)$.

We measure the CSD orientation only when at least 40 particles satisfy the CSD selection criteria, to avoid poorly sampled estimates. For CSD $k$, the apsidal angle is measured only when $e_k\geq0.02$, because the direction of the coherent eccentricity vector becomes poorly defined as its magnitude approaches zero. Samples failing either criterion are treated as undefined and are not interpolated.

\begin{figure*}
    \centering
    \includegraphics[width=\linewidth]{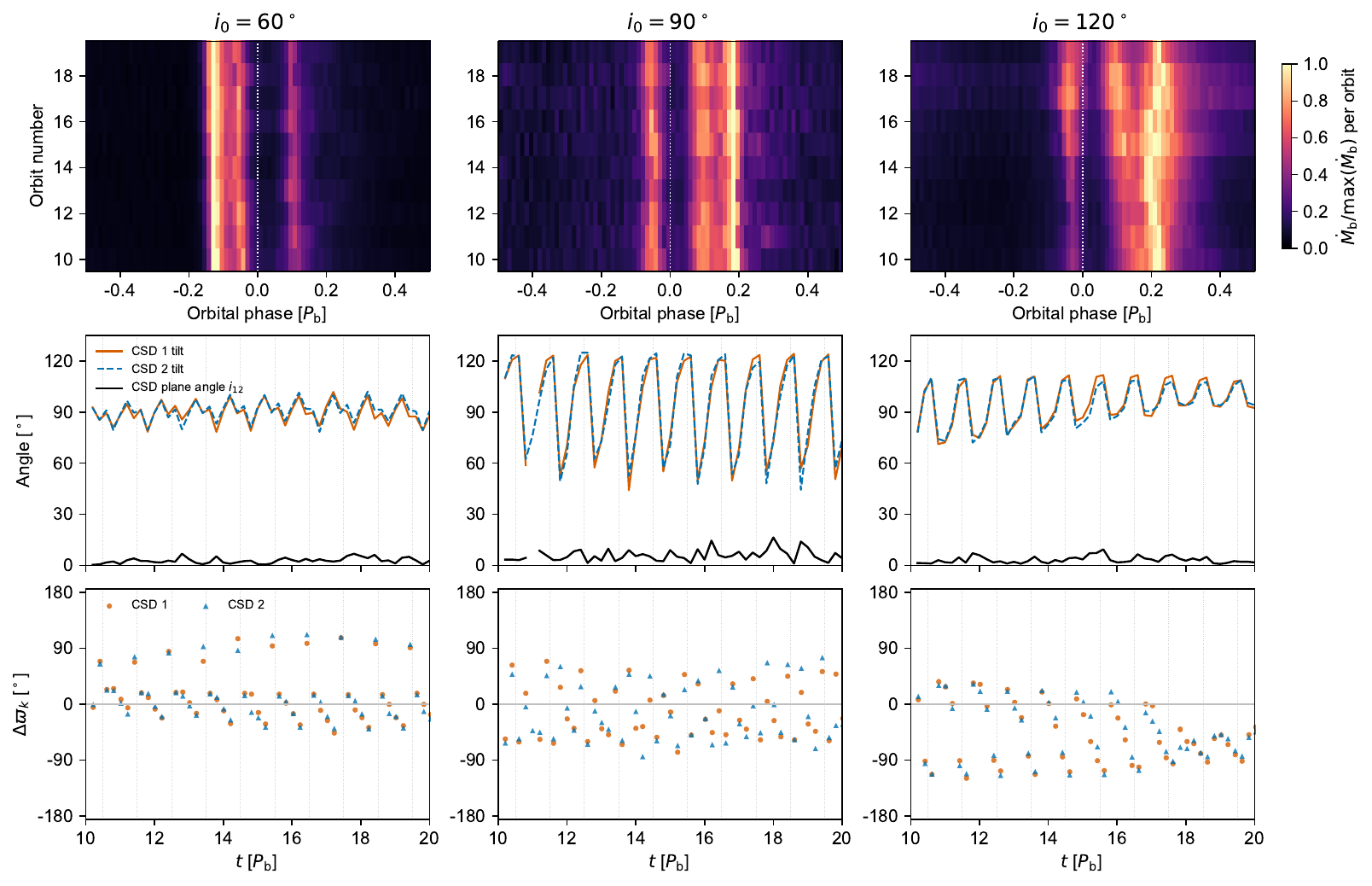}
    \caption{Accretion variability and CSD geometry across individual binary orbits in the $10$-$20\,P_\mathrm{b}$ interval. Columns correspond to initial CBD tilts of $60^\circ$, $90^\circ$, and $120^\circ$. Top: binary accretion rate as a function of orbital phase and orbit number, with each orbit normalized independently by its maximum to emphasize pulse timing rather than amplitude. Middle: the individual CSD tilts and the acute angle between the two CSD planes. Bottom: signed CSD apsidal angles, $\Delta\varpi_k$, where $k=1$ and $2$ refer to the circumprimary and circumsecondary disks, respectively. Each apsidal angle is measured relative to the periastron direction of the corresponding stellar barycentric orbit projected onto the instantaneous CSD plane. Vertical dotted lines mark periastron passages. CSD orientations are measured only when at least 40 particles satisfy the selection criteria; apsidal angles additionally require the coherent eccentricity magnitude of the corresponding CSD, $e_k$, to satisfy $e_k\geq0.02$. Samples that fail either criterion are left undefined and are not interpolated.}
    \label{fig:geometry_robustness}
\end{figure*}

Figure~\ref{fig:geometry_robustness} tests the robustness of the phase-averaged result using the individual binary orbits in the $10$-$20\,P_\mathrm{b}$ interval. Pulse amplitudes vary from one orbit to another, but the dominant pulse remains before periastron in the $i_0=60^\circ$ case and after periastron in the $i_0=90^\circ$ and $120^\circ$ cases. The individual CSD tilts also vary, while the acute angle between the two CSD planes generally remains small; the apsidal angles show stronger variations with phase and orbit number. Thus, phase averaging smooths these variations without changing the qualitative pre- versus post-periastron distinction among the three models. For the two CBD annuli defined in Section~\ref{sec:setup}, the coherent eccentricity magnitude remains $e_D<0.02$ throughout the same interval. We therefore do not assign a common apsidal direction to the CBD. This indicates that the gas does not maintain a sufficiently coherent common periastron; it does not imply that individual gas trajectories are circular.

\bibliography{CBD_short_term}{}
\bibliographystyle{aasjournalv7}


\end{CJK*}
\end{document}